\documentstyle[proceedings]{jd}
\input psfig
\input epsf
\def\etal{{\it et~al.\ }}
\def\aa #1 #2 {A\&A, #1, #2}
\def\aas #1 #2 {A\&AS, #1, #2}
\def\acm #1 #2 {ACM-Trans Math Software, #1, #2}
\def\ada #1 #2 {Ann Astrophys, #1, #2}
\def\agabstr #1 #2 {Astr Ges Abstr Ser, #1, #2}
\def\aj #1 #2 {AJ, #1, #2}
\def\anach #1 #2 {Astr Nachr, #1, #2}
\def\apj #1 #2 {ApJ, #1, #2}
\def\apjl #1 #2 {ApJL, #1, #2}
\def\apjs #1 #2 {ApJS, #1, #2}
\def\araa #1 #2 {ARAA, #1, #2}
\def\apss #1 #2 {ApSpaceS, #1, #2}
\def\celmech #1 #2 {Cel Mech, #1, #2}
\def\esom #1 #2 {ESO Messenger, #1, #2}
\def\fundcp #1 #2 {FunCosP, #1, #2}
\def\jcp #1 #2 {J Comp Phys, #1, #2}
\def\jfm #1 #2 {J Fluid Mech, #1, #2}
\def\jmp #1 #2 {J Math Phys, #1, #2}
\def\ma #1 #2 {Mitt Astr Ges, #1, #2}
\def\mn #1 #2 {MNRAS, #1, #2}
\def\nat #1 #2 {Nat, #1, #2}
\def\obs #1 #2 {Observatory, #1, #2}
\def\pasj #1 #2 {PASJ, #1, #2}
\def\pasp #1 #2 {PASP, #1, #2}
\def\phyr #1 #2 {PhysRep, #1, #2}
\def\physd #1 #2 {Physica D, #1, #2}
\def\rpp #1 #2 {RepProgPhys, #1, #2}
\def\ssr #1 #2 {Sp Sci Rev, #1, #2}
\def\iau127#1{in de Zeeuw P.T. ed, Structure and Dynamics of 
     Elliptical Galaxies, IAU Symp.~No.~127. Reidel, Dordrecht, p.~#1}

\def\in#1#2#3#4#5#6{in: #1%
\if#2-%
\else%
, #2%
\fi%
\if#3-%
\else%
, ed.\ #3%
\fi%
\if#5-%
 {\if#4-%
 \else,%
   (#4)%
 \fi}%
\else%
 {\if#4-%
, (#5)%
\else%
, (#5:#4)%
\fi}%
\fi%
\if#6-%
.%
\else%
, #6.%
\fi%
}

\def\spose#1{\hbox to 0pt{#1\hss}}
\def\lta{\mathrel{\spose{\lower 3pt\hbox{$\mathchar"218$}}
     \raise 2.0pt\hbox{$\mathchar"13C$}}}
\def\gta{\mathrel{\spose{\lower 3pt\hbox{$\mathchar"218$}}
     \raise 2.0pt\hbox{$\mathchar"13E$}}}
\def\equal{\! = \!}

\def\=#1{\overline{#1}}

\def\d{{\rm d}}

\def\deg{^\circ}             

\def\kms{{\rm\,km\,s^{-1}}}

\def\pc{{\rm\,pc}}
\def\kpc{{\rm\,kpc}}

\begin{opening}

\title{The Galactic Bar}

\author{Ortwin Gerhard, James Binney, HongSheng Zhao}

\institute{$^1$Astronomisches Institut, Universit\"at Basel, Switzerland \\
	   $^2$Department of Physics, Oxford University, UK \\
	   $^3$Sterrewacht, Leiden, The Netherlands \\
      gerhard@astro.unibas.ch, binney@thphys.ox.ac.uk, hsz@strw.LeidenUniv.nl}

\end{opening}

\runningtitle{The Galactic Bar}

\begin{document}

\begin{abstract}
\small
We summarize recent work on the structure and dynamics of
the Galactic bar and inner disk. Current work focusses on constructing
a quantitative model which integrates NIR photometry, source count
observations, gas kinematics, stellar dynamical observations, and
microlensing. Some avenues for future research are discussed.
\end{abstract}

\null\vskip-1cm

\section{Introduction}

There is substantial, and still growing, evidence for a bar in the
inner Galaxy. This includes the NIR light distribution, various source
counts, the atomic and molecular gas morphology and kinematics, and
probably the large optical depth to microlensing. See Gerhard (1996)
and Kuijken (1996) for recent reviews.

The existence of a rotating bar in the inner $\sim 3 \kpc$ of the
Milky Way is therefore no longer controversial. This is an important
development in that it changes the way in which we have to think about
the Galaxy's evolutionary history.

The emphasis of recent work has shifted towards determining parameters
like the orientation, size, and pattern speed of the bar, and towards
constructing a unifying quantitative model, within which the various
observational results can be coherently explained, and which predicts
the dynamical state of the Galactic bulge and inner disk.  The purpose
of this Joint Discussion Session is to present current work towards
this goal, and to discuss the necessary next steps for arriving at
such a model.

To make further progress, we must answer several questions. What are the
detailed shape, orientation, length, and pattern speed of the bar?
Should a distinction be made between the barred bulge and the bar in
the disk, between components of different age and metallicity?  Do the
NIR light, clump giants and IRAS sources trace the same component?
What is the mass distribution? Can we understand the kinematics of the
Galactic centre gas and the $3 \kpc$ arm? Can we predict the locations
of the Galactic spiral arms?  How does the large microlensing optical
depth fit in?  Do dynamical models predict the observed stellar
kinematics in Baade's window and the correct microlensing duration
distribution?  Many of these questions might be answered in the next
year or two.

\section{Photometric structure (Ortwin Gerhard)}

The currently best models for the distribution of old stars in the
inner Galaxy are based on the COBE/DIRBE NIR data. The COBE data have
complete sky coverage, and broad-band emission maps are available over
a large wavelength range from the NIR to the FIR. These aspects
outweigh the disadvantages of this dataset: the relatively low spatial
resolution, the residual effects of dust absorption, and the fact that
it contains no distance information. Because extinction is important
towards the Galactic nuclear bulge even at $2\,\mu$m, the first task
is to correct (or `clean') the DIRBE data for the effects of
extinction.  Arendt et al.\ (1994) did this by assuming that the dust
lies in a foreground screen, while Spergel, Malhotra \& Blitz (1997)
and recently Freudenreich (1997) used a fully three-dimensional model
of the dust distribution.

Dwek et al (1995) were the first to model the inner Galaxy's NIR
emissivity $j({\bf r})$ by matching parametric models to the cleaned
data of Arendt et al.  Recently, Binney, Gerhard \& Spergel (1997,
BGS) used a Richardson--Lucy algorithm to fit a non-parametric model
of $j({\bf r})$ to the cleaned data of Spergel et al.  This algorithm
has built in the assumption of eight-fold (triaxial) symmetry with
respect to three mutually orthogonal planes of arbitrary orientation
with respect to the Sun--Galactic centre line. It exploits the fact
that for such triaxially symmetric bars the perspective effects from
the location of the Sun contain information about the bar's shape and
orientation (Blitz \& Spergel 1991).  When the orientation of the
symmetry planes is fixed, the recovered emissivity $j({\bf r})$
appears to be essentially unique (Binney \& Gerhard 1996, Bissantz
\etal 1997), but physical models for the COBE bar can be found for
a range of bar orientation angles, $15\deg \lta \phi \lta 35\deg$
(BGS). $\phi$ measures the angle in the Galactic plane between the
bar's major axis at $l>0$ and the Sun--centre line.  Zhao (1997b)
has given an illustration of this non--uniqueness in terms of
the even part of the bulge density distribution (see Fig.~4 below).


Fig.~1 shows the resulting deprojected luminosity distribution for
$\phi=20\deg$. This shows an elongated bulge with axis ratios 10:6:4
and semi--major axis $\sim 2\kpc$, surrounded by an elliptical disk
that extends to $\sim 3.5\kpc$ on the major axis and $\sim 2\kpc$ on
the minor axis. There is a maximum in the NIR emissivity $\sim 3\kpc$
down the minor axis, which is probably due to incorrectly deprojected
strong spiral arms (see below), and which corresponds to the
ring--like structure discussed by Kent, Dame \& Fazio (1991). We
favour the $\phi=20\deg$ model because the dynamics of the large gas
velocities in the Galactic centre (Binney \etal 1991) and the
distribution of clump giants in the OGLE fields (Stanek \etal 1997)
point towards a near--end--on bar.

\begin{figure}
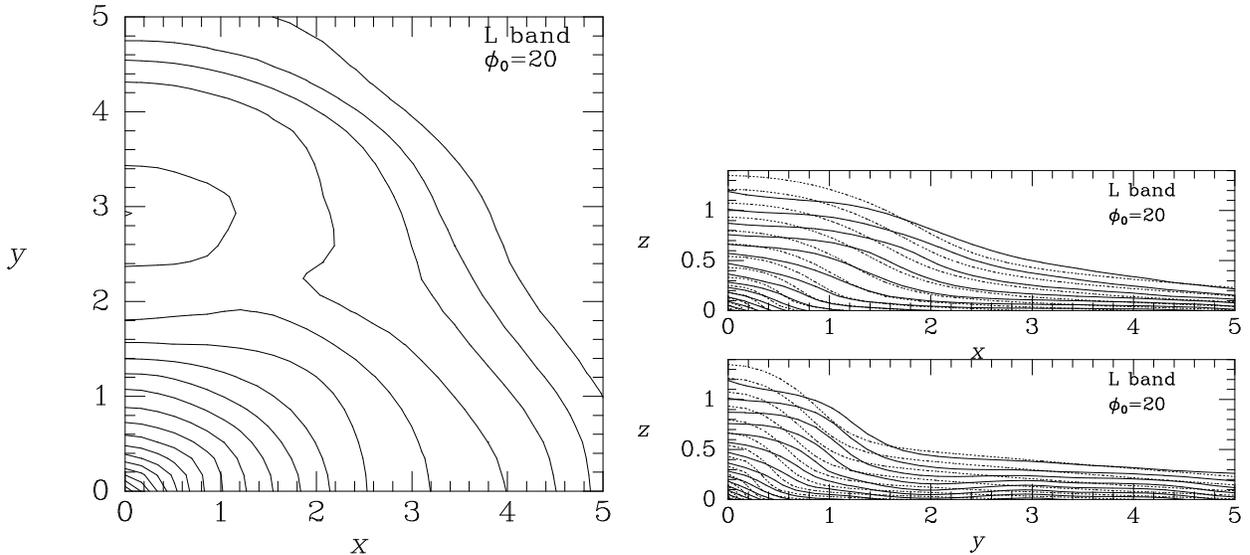

\hbox to \hsize{\psfig{figure=jd15_fig1a.ps,width=8cm}
	   \hfill \vbox{\psfig{figure=jd15_fig1b.ps,width=8cm} \par
			\psfig{figure=jd15_fig1c.ps,width=8cm} \vfill } }
\caption{Luminosity model for the inner Galaxy obtained by
  Lucy--Richardson deprojection of the cleaned COBE L--band data
  for $\phi=20\deg$. Left: Density projected along the $z$-axis, contours
  spaced by 0.1 dex. Right: Isodensity surfaces in the $zx$ and $zy$
  planes, contours spaced by 0.2 dex. Axis lengths in kpc. From BSG.}
\end{figure}

The largest point-source catalogue used for analysing the structure of
the Galactic bulge is that of the clump giant stars identified in the
OGLE colour-magitude diagrams (Stanek \etal 1997). These stars have a
small intrinsic luminosity spread ($\sim 0.2 - 0.3$ mag); thus their
apparent-magnitude distributions contain significant distance
information. They are visible throughout the inner Galaxy, and
reddening corrections can be made.  The drawback of the present
samples is that only a small number of lines--of--sight have been
observed.  Stanek \etal (1997) fitted a number of parametric models
from Dwek \etal (1995) to their data. Their best bulge model is one
with an exponential density distribution, axial ratios 10:4:3, and a
bar angle of $\phi=20 -30 \deg$. This model does not have a disk
component, whereas the COBE data imply that the foreground disk
contains $\sim 30\%$ of the light even in Baade's window.  The Stanek
\etal\ model must thus approximate the bulge {\it and\/} disk
luminosities along the 12 fitted lines--of--sight. Because the number
of lines--of--sight is small, and the fitted density models contain
only few parameters, systematic errors in the derived parameter
values are likely.

Nikolaev \& Weinberg (1997) have reanalyzed the IRAS variable
population.  This sample has a large spatial coverage near the plane
and distance information from the known range of AGB star
luminosities, but suffers from selection effects and incompleteness.
Reconstructing the density with a likelihood approach from about 5500
stars that satisfied their selection criteria, Nikolaev \& Weinberg
(1997) find a barred distribution (again, including the disk bar) with
length $\sim 3.5 \kpc$, axial ratio in the plane 10:4, and orientation
$\sim 21 \deg$.

While this combined work shows that the bar angle $\phi$ is probably
close to $\phi=20\deg$, there is less agreement in the axis ratios of
the bar. The work of BGS shows that the disk contribution in the
central few $\kpc$ is large. A simple bulge--disk decomposition will
not be possible. Clearly then, the next step is to combine the area
coverage of the COBE data with the distance information in the clump
giant samples, and perhaps later add the IRAS source population, in a
non-parametric approach. Work along these lines has begun.  This may
constrain the shape parameters and the orientation of the bar much
more tightly if all works out, or it may tell us that different
tracers of the Galactic bulge and inner disk represent components with
different spatial distributions. Both would be interesting.

\section{Gas dynamics (Ortwin Gerhard)}

It has long been suspected that the non--circular and forbidden
velocities in the HI and CO $(l,v)$--diagrams are caused by a
non--axisymmetric component of the gravitational potential in the
inner Galaxy (for references see Gerhard 1996). With quantitative
models of the old stellar mass distribution now becoming available,
we can make detailed models of the Galactic gas flow, and ask
whether prominent structures in the observed $(l,v)$--diagrams
such as the $3\kpc$ arm, the $4\kpc$ ring, the Galactic centre
parallelogram, etc, can be obtained.

In a first attempt Englmaier \& Gerhard (1997) have used the
luminosity distribution derived by BGS from the cleaned COBE L--band
data of Spergel \etal (1997) for various $\phi=10-30\deg$.  With the
assumption that the NIR light traces mass, i.e., constant $M/L_L$,
models for the mass distribution and gravitational potential of the
inner Galaxy were constructed. The main uncertainties in these models
concern the inner disk, because the details of the in--plane dust
distribution are uncertain, and because the Richardson--Lucy
deprojection algorithm is unable to reliably deproject spiral arms
(BGS).

In the resulting gravitational potentials, quasi--equilibrium gas flow
models were computed, using an effective 20000 SPH particles in two
dimensions, and assuming point symmetry with respect to the Galactic
centre. The most important parameter to be determined is the pattern
speed; this is restricted by the requirement that the $3 \kpc$ arm
must lie inside corotation and the $4 \kpc$ ring (actually, a pair of
spiral arms) at or beyond corotation. The best current models have
$R_{\rm corot} = 3.4 \kpc$ for a pattern speed $\Omega_p=55 {\rm
km/s/kpc}$.

These models clearly show a four--armed spiral structure outside
corotation, as indicated by observations (Vall\'ee 1995). They also
show that the luminosity concentrations along the bar's minor axis
at $\sim 3\kpc$ central distance must be at least partially real;
removing these from the mass model results in an unstructured gas disk
outside $4 \kpc$, rather unlike observations. Most likely, these
concentrations represent the NIR light from stellar spiral arms
near--coincident with the gas arms. Inside corotation the $3 \kpc$ arm
is one of the arms emanating from the nearby end of the bar.  The
transition between $x_1$ and $x_2$ orbit flow in these models is
highly dynamical; thus in these models, the molecular parallelogram is
non--stationary and not well--resolved.

The current best model includes a dark component with asympototic
circular velocity $v_c=200\kms$. With a galactocentric solar radius of
$R_0=8\kpc$, a bar angle and corotation radius of $\phi =20\deg$ and
$R_{\rm corot} = 3.4 \kpc$, this does a rather good job at reproducing
the observed spiral arm tangents as seen from the Sun (Fig.~2). The
bar angle is not very well--constrained; $\phi =25\deg$ is equally
possible.  From fitting grid--based hydrodynamic models in analytic
Galaxy potentials to the low--column density contours in the HI
$(l,v)$--diagram, Weiner \& Sellwood (1997) favour bar angles around
$\sim 30\deg$. The low--density regions are not well--resolved in the
SPH models, so no direct comparison is currently possible.

Fux (1997) has computed a large number of combined N-body and SPH
models from a grid of initial conditions, and has fitted these to the
COBE NIR surface brightness contours and the observed
$(l,v)$--diagrams. These models appear to be similar to those of
Englmaier \& Gerhard (1997) in several respects; they too reproduce a
number of features in the observed gas morphology and kinematics.
While they cannot be exactly made--to--measure, they include effects
from asymmetries (well--known to be of some importance in the Galactic
centre molecular gas) and are not necessarily stationary.

With this recent work we may be close to understanding the gross
structure of the inner Galactic gas flow. Comparing the results from
the two approaches, ab--initio versus
model--inferred--from--observations, and comparing results obtained
with different hydrodynamical techniques, will be particularly helpful
in isolating the model--independent results.

\begin{figure}
\hbox to \hsize{\psfig{figure=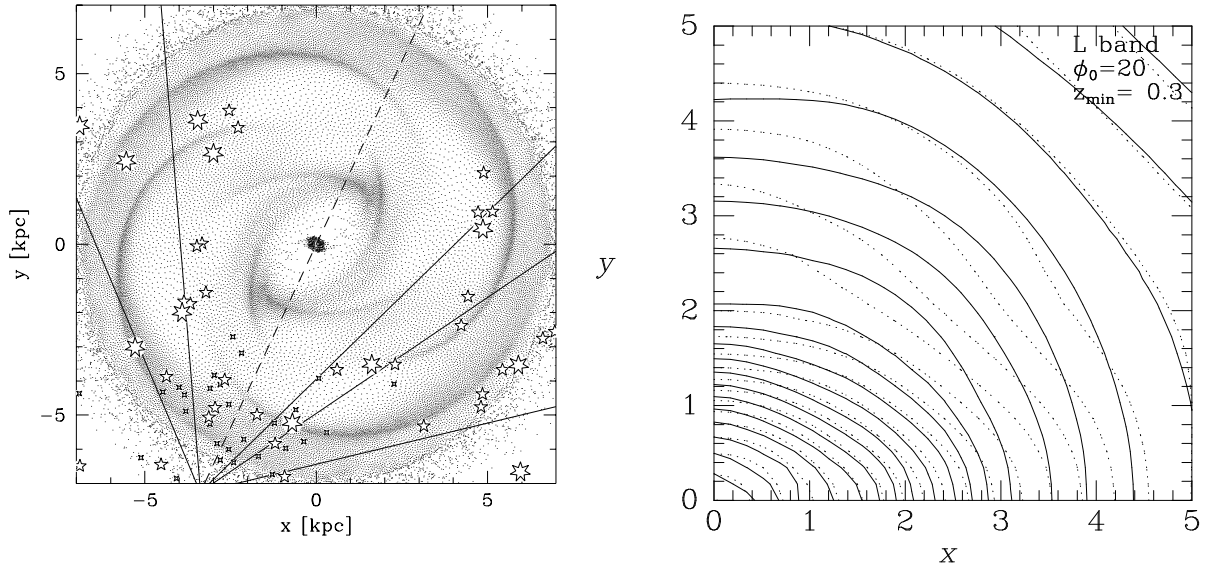,width=8cm}
	   \hfill \psfig{figure=jd15_fig3.ps,width=8cm} }
\caption{(Left panel) Gas flow in the inner Galaxy. The SPH particle
     distribution is plotted to trace the spiral arm shocks. The solid
     lines from the position of the Sun denote tangent point
     directions of Galactic spiral arms, located approximately at
     $l\equal 48\deg$, $l\equal 30\deg$, $l\equal -21\deg$ (the `$3
     \kpc$ arm'), $l\equal -31\deg$, $l\equal -52\deg$. From
     Englmaier \& Gerhard (1997). Stars denote the positions of HII
     regions from Georgelin \& Georgelin (1976). }
\caption{(Right panel) Two
     models of the inner Galaxy projected from $z=300\pc$ vertically
     upwards. The dashed contours are for the BGS photometric model
     while the full contours are for a preliminary dynamical model
     that was designed to fit the BGS model.}
\end{figure}

\section{Dynamical Models of the Galactic Bulge (James Binney)}

While the inner Galaxy is now generally agreed to be barred, the
orientation, length and pattern speed of this bar remain controversial.
Since dynamics must connect these parameters, the obvious next step is to
upgrade photometric models of the inner Galaxy to fully dynamical models.

We believe that the photometric model of BGS is the best available
because (i) it is based on COBE/DIRBE data cleaned with a fully
three-dimensional model of the dust distribution, and (ii) because
their non-parametric technique fits the cleaned data more exactly than
does the previous parametric approach of Dwek et al.~(1995).  Zhao
(1996) matched a dynamical model to the photometric model of Dwek et
al. In Oxford we have used an extended version of the fitting
technique introduced by Zhao to do the same thing for the photometric
model of BGS.  This work is still in progress (H\"afner et al.\ in
preparation) so we show only a few preliminary results, concentrating
on our methodological innovations.

Like Zhao we start by choosing a pattern speed $\Omega_p$ and use an
enhanced version of Schwarzschild's technique. Hence, we calculate
orbits in an assumed potential and then populate them so that the sum of the
densities contributed by every orbit in our orbit library reproduces the
originally assumed density distribution.

Schwarzschild was primarily concerned with the case in which all
relevant orbits are regular -- that is, have three effective isolating
integrals. Much of the phase space of a rapidly rotating bar is not
occupied by such orbits, but by stochastic orbits. Stochastic orbits
are liable to evolve on long timescales. Consequently, they can be
included in the orbit library only after they have been followed for a
Hubble time. When they are followed for so long, one finds that all
stochastic orbits of a given energy closely resemble one another. This
being so one can eliminate stochastic orbits from the library
completely if one (a) includes for each energy a `composite orbit' and
(b) permits the weights of regular orbits to become negative in
certain cases.

The concept of a composite orbit was introduced by Zhao (1996). It is the
phase-space distribution that is produced by a distribution function $f$
that is a $\delta$-function in the Hamiltonian $H$. A stochastic orbit is
formed by subtracting from the composite orbit of the same energy
appropriate amounts of every regular orbit of that energy. Hence we can
replace stochastic orbits with composite orbits provided we permit the
weights of regular orbits to be negative. The only snag is that we must not
permit these weights to be so negative that the overall phase-space density
becomes negative anywhere in phase space.

While Zhao replaced stochastic orbits with composite orbits, he did
not allow weights to be negative because he had no way of determining
the overall phase-space density $f$. Our technique for determining $f$
is as follows.  We choose a sampling density $F(\b r,\b
p)=\sum_{i=1}^3\rho_i(\b r)h_i(\b p)$ that we believe lies close to
the final phase-space density in regular regions of phase space. $F$
is normalized such that its integral throughout phase space evaluates
to unity. Then each orbit in the library is launched from the
phase-space volume $\d^3\b r\d^3\b p$ with probability $F\,\d^3\b
r\d^3\b p$, and, as the orbit is integrated, we evaluate the time average
$\={F}$ of $F$ along the orbit. If a given orbit is subsequently
assigned weight $w$, one may show that along this orbit the final
phase-space density in the model is $f=w\={F}+w_c$, where $w_c$ is the
weight assigned to the composite orbit of the same energy.

We used Liapunov exponents to select $\sim20\,000$ regular orbits from
$\sim250\,000$ initial conditions in the potential that one obtains by
assigning some mass-to-light ratio $\Upsilon$ to the photometric model of
Binney et al. A Richardson--Lucy algorithm was then used to assign weights
to both these orbits and 2000 composite orbits. 

As we saw above, the BGS photometric model contains structures at
$|z|\lta300\,$pc that are probably incorrectly deprojected spiral
arms. We would not expect to be able to reproduce such structures with
the present technique. Therefore in Fig.~3 we
compare the projection from $z=300\,$pc upwards for (a) our final
dynamical model (full contours) and (b) the BGS photometric model
(dashed contours). While the two contour sets resemble each other in
many respects, they do not agree precisely. This disagreement is not
unexpected because the original mapping from the COBE data to a
photometric model is surely not unique and one does not expect to be
able to model an arbitrary density distribution dynamically. Therefore
our next step will be to calculate orbits in the potential of our
dynamical model and to populate them so as to optimally reproduce the
COBE brightness distribution {\em on the sky}. We anticipate obtaining
a good fit, which we shall be able to improve by calculating orbits in
the potential of the new dynamical model and again fitting to the COBE
data.

\section{Microlenses towards the Galactic Bulge/Bar (HongSheng Zhao)}

Galactic microlensing started out as a unique technique of detecting
the Galactic dark matter when in the form of compact baryonic dark
objects (such as brown dwarfs in the Galactic disc, and Machos in the
Galactic halo) by obtaining light curves of $10^{6-7}$ of stars
towards the LMC/SMC and the Galactic bulge and searching for
occasional transient gravitational amplification of light by these
dark objects (see review of Paczy\'nski 1996).  While more than 200
microlenses towards the bulge and about a dozen towards the LMC have
been detected so far (Udalski et al. 1994, Alcock et al. 1997a,b), it
is still unclear whether they are the baryonic dark matter which we
have been searching for or simply ordinary faint, low-mass stars.
Towards the LMC, some non-conventional models predict that the
Magellanic Clouds are shrouded with a common faint stellar halo
created by tidal interaction of the binary in their recent close
encounter and these stellar tidal debris in the near/far side can
significantly microlens or be microlensed by stars in the LMC (Zhao
1997a and references therein) with an optical depth comparable to the
observed value of $\tau_{\rm obs} \sim (3 \pm 1) \times 10^{-7}$
(Alcock et al. 1997b).  These predictions need to be tested
observationally before one can draw conclusions of any Machos in the
halo, and the ratio of baryonic to non-baryonic dark matter.  Towards
the Galactic bulge/bar, microlensing by low-mass stars in the near
side of the bar is expected to dominate the lenses in the disc (Zhao,
Spergel \& Rich 1995), and brown dwarfs should not make up most of the
mass in the bar and the disc (Zhao, Rich \& Spergel 1996).  However,
as I will point out, previous models which try to bring together the
microlensing data with the COBE/DIRBE maps of the bar still suffer two
serious problems.

First, we have seen that the 3D distribution of the bar cannot be
uniquely recovered from the COBE/DIRBE maps (BGS, Zhao 1997b).
Moreover, with Schwarzschild's method one can probably generate
photometrically acceptable and dynamically consistent bars with a
range of different pattern speeds and streaming-velocity fields.
Consequently, the potential of the bar and the velocity distribution
of stars in the bar will be non-unique (cf Fig.~4).  This
non-uniqueness propagates to the predicted lens masses $m$ because $m
\propto \left(\mu_s-\mu_l \right)^2 t_{\rm obs}^2 {D_l D_s \over
D_s-D_l}$, where $\mu_l$ and $D_l$ are the proper motion and distance
of the lens, and $\mu_s$ and $D_s$ are that of the source.  For the
same data of the event duration distribution, the predicted lens
masses could differ in some cases by as much as a factor of 2 in
different models of the bar (Zhao \& de Zeeuw 1997).  This renders it
meaningless to derive details of the mass spectrum using lensing data
without first lifting the degeneracy in the dynamical models.

\begin{figure}
\centerline{\epsfysize=7cm\epsfbox{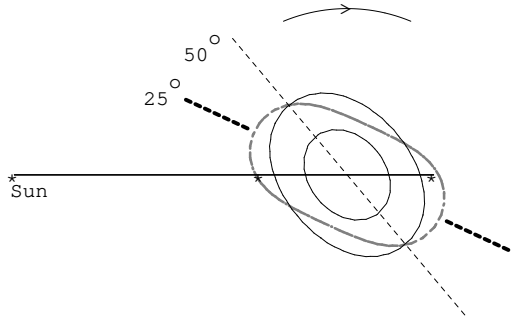}}
\vskip-1.0cm
\caption{Equatorial slices through two triaxial bar models that have
  identical surface brightness maps from the Sun's perspective: an
  extended boxy bar at $\alpha=25^\circ$ (thick dashed contours) and a
  more centrally concentrated and rounder bar at $\alpha=50^\circ$
  (thin solid contours).  The former should be more efficient in
  microlensing than the latter as the optimal geometry for
  microlensing is a nearly homogeneous and rectangular bar oriented
  with the line of sight passing through its diagonal line (Zhao \&
  Mao 1996); such a geometry would put a high density of lenses well
  in front of the sources, hence increase the optical depth $\tau
  \propto \rho (D_s-D_l)^2$.  Moreover, the boxy, less centrally
  concentrated bar also predicts a shallower potential well hence less
  massive lenses because $m$ decreases with decreasing velocity and
  increasing size of the bar.  Bars with the same 3D density
  distribution but different pattern speed are yet another source of
  non-uniqueness.}
\end{figure}

Second, current theoretical models of the bar and the disc cannot
fully account for the very high optical depth observed towards the
bulge.  As shown in Zhao \& Mao (1996), even when the geometry of the
bar is artificially tailored to enhance microlensing (cf.~Fig.~4), the
maximum optical depth is still lower than the observed by at least
$1\sigma$.  More physical bar models that fit either the star-count
data for bulge red clump giants (Stanek et al.\ 1997) or the
COBE/DIRBE infrared surface-brightness distribution (Bissantz et al.\
1997), yield $\tau \sim (0.8-1.2) \times 10^{-6}$, which is about
$(2-3)\sigma$ smaller than the value determined from observations of
bulge red-clump giants (Alcock et al.\ 1997a) $\tau_{\rm obs} \sim (4
\pm 1.2) \times 10^{-6}$.

While there is every prospect of breaking the degeneracy in the first
problem by integrating microlensing data with star-count data for
bulge red-clump giants and data for the gas/stellar kinematics, the
implied extra lenses of the second problem are puzzling.  Models that
go beyond the standard COBE bar-plus-disc model would seem inevitable
if the high optical depth derived by the experimental teams holds up
over the next years as the number of microlensing events increases and
our understanding of the experimental efficiency improves.

{}

\end{document}